\documentclass[12pt]{article}

\usepackage{amsmath}
\usepackage{amssymb}
\usepackage{mathtools}
\newcommand{\be}{\begin{equation}}
\newcommand{\bea}{\begin{eqnarray}}
\newcommand{\ee}{\end{equation}}
\newcommand{\eea}{\end{eqnarray}}

\begin{document}

\setcounter{page}0
\renewcommand{\thefootnote}{\fnsymbol{footnote}}
\begin{titlepage}
%
\vskip .7in
\begin{center}
{\Large \bf Lagrangian formulation of massive fermionic higher spin fields \\
on a constant electromagnetic background
 } \vskip .7in {\large I.L. Buchbinder$^{ab}$\footnote{e-mail: {\tt joseph@tspu.edu.ru}},
V.A.Krykhtin$^{ac}$\footnote{e-mail: {\tt  krykhtin@tspu.edu.ru }} and
 M.Tsulaia$^d$\footnote{e-mail: {\tt  mirian.tsulaia@canberra.edu.au}  }}
\vskip .2in {$^a$ \it Department of Theoretical Physics, Tomsk State Pedagogical University,\\
Tomsk, 634061, Russia} \\
\vskip .2in
{\it ${}^b$National Research Tomsk State University, Russia}\\
\vskip .2in {$^c$ \it
Laboratory of Mathematical Physics, Tomsk Polytechnic University, \\
Tomsk, 634050, Russia} \\
\vskip .2in { $^d$ \it Faculty of  Education, Science, Technology and Mathematics, \\
University of Canberra, Bruce ACT 2617, Australia}

\begin{abstract}
We consider  massive half-integer higher spin fields coupled to an
external constant electromagnetic field in flat space of an arbitrary
dimension and construct a gauge invariant Lagrangian in the linear
approximation in the external field. A procedure for finding the
gauge-invariant Lagrangians is based on the BRST construction where
no off-shell constraints on the fields and on the gauge parameters are
imposed from the very beginning. As an example of the general
procedure, we derive a gauge invariant Lagrangian for a massive
fermionic field with spin 3/2  which contains  a set of
auxiliary fields and gauge symmetries.
\end{abstract}

\end{center}

\vfill

\end{titlepage}
\renewcommand{\thefootnote}{\arabic{footnote}}
\setcounter{footnote}{0}

\section{Introduction}

Despite   the recent progress in higher spin gauge theories (see
e.g.  \cite{Vasiliev:1999ba}-- \cite{Gomez:2014dwa} for review of
various aspects of the subject) there are still a number of problems to
address. Construction of the interacting Lagrangians of massive
higher spin fields on various
backgrounds and study of the properties of these systems is one of these
problems. Apart from being interesting in its own right, it is also
important from the string theory perspective \cite{Porrati:2010hm}.
As  is well known, string theory contains an infinite tower of
massive higher spin modes and therefore it is important to
understand on which backgrounds these fields can propagate
consistently.

Although many aspects of Lagrangian formulation of free fermionic
higher spin fields have been studied well enough (see e.g.
\cite{Buchbinder:2006nu}-- \cite{Buchbinder:2007vq} and references
therein) the problem of interacting fermionic fields is much less
understood than the problem of interacting bosonic fields
(see also \cite{Gomez:2014dwa} for a recent review). In particular,
that the cubic vertices which include fermionic higher spin fields
have been constructed in the light cone framework in
\cite{Metsaev:2007rn} and various problems of interaction with
gravitational and electromagnetic fields have been addressed in
\cite{Klishevich:1998yt}--\cite{Buchbinder:2014ata}\footnote{Also
one points out the papers \cite{Rahman:2013}, \cite{Rahman:2014}
where non-Lagrangian equations of motion for higher spin fields in
the external fileds have been considered.}.

When considering interactions of massive fields with spin more than
zero with a nontrivial background one faces several difficulties
such as  superluminal propagation and
violation of
the number of physical degrees of freedom. The requirement that no
superluminal propagation takes place imposes in general certain
conditions on the background fields
\cite{Velo:1969bt}--\cite{Velo:1970ur} (see also \cite{BKL} for a recent discussion). Similarly, when turning on nonzero  background
fields the invariance of the initial system under its gauge
transformations can be partially or completely lost and this means
in turn that nonphysical polarizations can appear in the spectrum.
The requirement of preserving  of physical degrees of freedom
generically imposes some extra conditions on the background. The
question is therefore to find if a background under consideration is
physically acceptable i.e., if it satisfies the constraints imposed by the
above mentioned conditions.

In this paper we consider a problem of interaction of massive
totally symmetric fermionic higher  spin fields with constant
electromagnetic (EM) background in Minkowski space  of an arbitrary
dimension $d$. These higher spin fields are described by
tensor--spinors with one spinorial index and an arbitrary number
$n=s-1/2$ of totally symmetric tensorial indices.
Our main aim is to derive the gauge invariant Lagrangian using the
method of BRST construction in the linear approximation in strength
$F_{\mu\nu}$ of the external field. This method in fact yields a
gauge invariant Lagrangian description for massive higher spin
fields in extended Fock space and therefore the Lagrangian will
contain, apart from the basic fields, some extra auxiliary fields such
as St\"{u}ckelberg fields. Some of these fields are eliminated with
the help of gauge transformations, some of the others should be
eliminated as a result of the equations of motion. Therefore, in
order to have a consistent gauge invariant description for massive
higher spin fields, one should have enough gauge freedom and have the
``correct" equations of motion, which ensure the absence of
ghosts\footnote{ One way to check this is to perform a complete
gauge fixing in the equations of motion and obtain the equations in
terms of basic fields. As a result one obtains equations defining
the spectrum of the theory and check if it is ghost free or not. }.
 Performing this analysis in a way similar to how it has been done
in \cite{Buchbinder:2013uha} one can  show that the preservation of
physical degrees of freedom indeed takes place for the Lagrangian
under consideration, provided that the terms containing the strength
of the external space are considered as a perturbation. Where the
problem of superluminal propagation of higher spin fields is
concerned we note that in the linear in $F_{\mu\nu}$ approximation
this problem does not arise at all due to antisymmetry of
$F_{\mu\nu}$ (see e.g.   \cite{Velo:1969bt} for a spin $3/2$ field).

The paper is organized as follows.  Section~\ref{Fne0} contains
our main results. After a brief reminder of construction of
Lagrangians for free massive fermionic higher spin fields we
introduce interaction with background electromagnetic fields by
modifying the operators which define the BRST charge. The
requirement that the modified operators form a closed algebra
determines free parameters which are present in the definition of
the operators. Then we present the corresponding BRST charge,
construct the Lagrangian and use a part of the BRST gauge
transformations to gauge away an infinite number of neutral bosonic
ghost variables from the Lagrangian. The remaining components of the
basic fields obey the Lagrangian field equations and these equations
still posses necessary  gauge invariance. Integrating the field
equations back into a Lagrangian we complete the construction of
gauge invariant Lagrangian and equations of motion in terms of a basic
massive fermionic higher spin field and appropriate auxiliary fields
interacting with a constant EM background.

 Section~\ref{secquartet} contains a more generic description in
terms of so called ``quartet formulation"
\cite{Buchbinder:2007ak}--\cite{Buchbinder:2008ss} (see also
\cite{Pashnev:1989gm}). This formulation is obtained from the one
given in  Section \ref{Fne0} by further use of the BRST gauge
transformations to gauge away some auxiliary fields which are
originally present in the system. In this way the Lagrangian
contains only one physical field and six auxiliary fields three of
which are Lagrangian multipliers. Let us note that in both cases the
fields and the parameters of gauge transformations do not contain
any off-shell conditions,  unlike the formulation of
\cite{Fang:1978wz}.

In Section ~\ref{ex-32} we give  a  description   of the
simplest example of the spin $\frac{3}{2}$ field interacting with a
constant EM background.

The final Section contains our conclusions and a discussion of
some open problems.

\section{Construction of gauge invariant Lagrangians} \label{Fne0}
\setcounter{equation}0
Let us briefly summarize the features of the BRST approach for the
construction of the gauge invariant free and interacting Lagrangians
(see \cite{Fotopoulos:2008ka} for a review). First one introduces a
set of operators that define a spectrum of the theory\footnote{In
free theory the spectrum is given with the help of the relations
defining either reducible or  irreducible representations of the
Poincare or AdS group.}. Provided these operators form a closed
algebra one builds a nilpotent BRST charge $Q$, which in turn yields
to a quadratic gauge invariant Lagrangian of the form
\begin{equation}
{\cal  L}  \sim \langle \chi |Q| \chi \rangle
\end{equation}
where $| \chi \rangle$ is a vector in an extended Fock space. The
gauge invariance of the Lagrangian under the linear gauge
transformations
\begin{equation}\label{L}
 \delta| \chi \rangle = Q| \Lambda \rangle
\end{equation}
is guaranteed by the nilpotency of the BRST charge $Q^2=0$. This
procedure is however slightly modified for the case of fermionic
higher spin fields, since the condition of the BRST invariance
\begin{equation} \label{Q}
Q|\chi\rangle=0
\end{equation}
cannot be integrated back into a Lagrangian in a straightforward
way. Rather one uses a part of  the gauge transformations (\ref{L})
to gauge away a part of the  auxiliary fields which are contained in
$|\chi \rangle$. The resulting field equations turn out to be
Lagrangian ones and they still possess enough gauge invariance to remove
all nonphysical polarizations (see
\cite{Buchbinder:2004gp} for the details).

The situation is even  more complicated if the closure of the
algebra of the initial set of operators requires inclusion of
certain additional operators into the system. These extra operators
can impose too strong conditions on the field $|\chi \rangle$ so
that there will be no nonzero solution to the equation (\ref{Q}). A
way out of this problem is the following (see
\cite{Buchbinder:2006nu}
\cite{Buchbinder:2004gp},\cite{Buchbinder:2007vq} for  fermionic
higher spin fields and \cite{Fotopoulos:2008ka} for a detailed
review of the BRST formulation for higher spin fields). One
introduces additional sets of  oscillator variables and builds
auxiliary representation of the generators of the algebra (i.e. of
the operators under consideration) in terms of these new variables.
Then one defines a modified set of operators as a sum of new and old
ones and therefore considers the problem in  an extended Fock space.
After that one builds BRST charge for modified generators in the
standard way since the generators form a closed algebra. It allows
us to construct a Lagrangian of the base of the BRST charge under
consideration.

After this reminder let us turn  to a description of massive
fermionic  higher spin fields. To this end we introduce the  Fock
space spanned by the oscillators
\begin{equation}
[a_\mu,a_\nu^+]=\eta_{\mu\nu},, \quad \eta_{\mu\nu}= diag (-1,1,...,1)
\end{equation}
and consider the  operators
\begin{equation} \label{properators}
t_0^\prime=i\tilde{\gamma}^\mu\partial_\mu,  \quad l^\prime_0=\partial^2-m^2, \quad l^\prime_1=
ia^\mu\partial_\mu, \quad t^\prime_1=\tilde{\gamma}^\mu a_\mu,
\quad  l_2^\prime=\tfrac{1}{2}a^{\mu}a_\mu,
\end{equation}
Here we introduce Grassmann odd ``gamma-marix like objects''
$\tilde{\gamma}^\mu$ and $\tilde{\gamma}$ which are connected with
the usual Grassmann even gamma-matrices $\gamma^\mu$ by relation
\cite{Buchbinder:2006nu}
\begin{equation}
\gamma^\mu=\tilde{\gamma}^\mu\tilde{\gamma},
\quad
\{\tilde{\gamma}^\mu,\tilde{\gamma}^\nu\}=-2\eta^{\mu\nu},
\quad
\{\tilde{\gamma}^\mu,\tilde{\gamma}\}=0,
\quad
\{\tilde{\gamma},\tilde{\gamma}\}=-2.
\end{equation}
The first of the operators in (\ref{properators}) corresponds to the
Dirac operator for the massive fermion, the second operator is the
d'Alembertian for a massive field, the third one is a divergence
operator, the fourth one is an operator which takes  a gamma-trace
and the fifth one is an operator which takes a trace. In order to
have a hermitian BRST charge we also introduce  operators which
are hermitian conjugate to the operators $ l^\prime_1, t^\prime_1$
and $ l_2^\prime$
\begin{equation} 
  l^{\prime +}_1=ia^{+\mu }\partial_\mu, \quad t^{\prime +}_1=\tilde{\gamma}^\mu a^+_\mu,
\quad  l_2^{\prime +}=\tfrac{1}{2}a^{+\mu }a^+_\mu,
\end{equation}
Finally in order to close the algebra one introduces the extra
operators
\begin{equation}
 g'_0=a^{+ \mu} a_\mu+\frac{d}{2}
\end{equation}
and $g'_{m}=m^2$. The operator $g'_0$ is a "particle" number
operator and its eigenvalues are  always strictly positive.
Therefore, we have a situation described earlier in this Section. We
introduce three sets of additional oscillator variables: two sets of
bosonic oscillator variables with commutation relations
\begin{equation} \label{b12}
[b_1,b_1^+]=1, \quad
[b_2,b_2^+]=1,
\end{equation}
and one set of fermionic oscillator variables
\begin{equation}
\{f,f^+\}=1. \label{f1}
\end{equation}
Using these new variables one can build auxiliary representation for
the original operators and define modified operators as
\cite{Buchbinder:2006nu}
\begin{align}
\label{op-0}
&
t_0=i\tilde{\gamma}^\mu\partial_\mu-\tilde{\gamma}m
&&
l_0=\partial^2-m^2
\\
&
l_1=ia^\mu\partial_\mu+mb_1
&&
l_1^+=ia^{+\mu}\partial_\mu+mb_1^+
\\
\label{op-t1}
&
t_1=\tilde{\gamma}^\mu a_\mu-\tilde{\gamma}b_1
+f^+b_2-2(b_2^+b_2+h)f
&&
t_1^+= a_\mu^+\tilde{\gamma}^\mu-\tilde{\gamma}b_1^++f^+-2b_2^+f
\\
&
\label{op-l2}
l_2=\tfrac{1}{2}a^{\mu}a_\mu+\tfrac{1}{2}b_1^2
+(b_2^+b_2+f^+f+h)b_2
&&
l_2^+=\tfrac{1}{2}a^{+\mu}a_\mu^++\tfrac{1}{2}b_1^{+2}+b_2^+
\\
& g_0=a^+_\mu a^\mu+b_1^+b_1+2b_2^+b_2+f^+f+\tfrac{d+1}{2}+h
&&
g_{m}=0
\label{op-g0}
\end{align}
where $h$ is an arbitrary real constant. The algebra of these
operators is given by Table 1.

In order to introduce an interaction of the fermionic fields with an
external constant EM background  field $F_{\mu\nu}=const$ we shall
proceed as follows. First we replace all the partial derivatives by
the $U(1)$ covariant ones $D_\mu=\partial_\mu-ieA_\mu$ and include
into the expressions of the operators\footnote{We shall denote these
new operators by the corresponding capital letters.}
(\ref{op-0})--(\ref{op-g0})  terms which  vanish in the limit
$F_{\mu\nu}\to0$. After that we require that the new  operators form
a closed algebra.

Before writing an ansatz for the operators let us note that since
the trace of a field and its traceless part are independent from
each  other one can shift the trace of a field so that the traceless
condition remains unchanged. Thus we suppose that the operators
related with the traceless condition $t_1$, $t_1^+$, $l_2$, $l_2^+$
as well as the number operator $g_0$  remain unchanged
\begin{align}
&T_1=t_1, &&T_1^+=t_1^+, &&L_2=l_2, &&L_2^+=l_2^+&&G_0=g_0.
\label{T1}
\end{align}
Moreover, since the oscillator variables $b_2$, $b_2^+$, $f$, $f^+$
(\ref{b12})--(\ref{f1}) are included only in operators (\ref{T1})
(see also the expressions (\ref{op-t1})--(\ref{op-g0})) we  assume
that these variables  are not present in the expressions of the
operators $T_0$, $L_0$, $L_1$, $L_1^+$.

Since we are going to consider only the linear in $F_{\mu\nu}$ approximation we take the following ansatz for the operators
\begin{eqnarray}
L_1&=&ia^\alpha D_\alpha+mb_1
+a^\alpha F_{\alpha\sigma}D^\sigma \sum_{k=0}^\infty f_{0k}\,b_1^{+k}b_1^k
+\tilde{\gamma}\tilde{\gamma}^\tau F_{\tau\sigma}D^\sigma \sum_{k=0}^\infty f_{2k}\,b_1^{+k}b_1^{k+1}
\nonumber
\\
&&{}
+a^{+\mu} F_{\mu\sigma}D^\sigma \sum_{k=0}^\infty f_{4k}\,b_1^{+k}b_1^{k+2}
+\tilde{\gamma}^{\mu\nu}F_{\mu\nu} \sum_{k=0}^\infty d_{0k}\, b_1^{+k}b_1^{k+1}
\nonumber
\\
&&{}
+\tilde{\gamma}\tilde{\gamma}^{\sigma}F_{\sigma\alpha}a^\alpha\sum_{k=0}^\infty d_{2k}\,b_1^{+k}b_1^k
\nonumber
\\
&&{}
+a^{+\mu}F_{\mu\alpha}a^\alpha \sum_{k=0}^\infty d_{8k}\, b_1^{+k}b_1^{k+1}
+\tilde{\gamma}\tilde{\gamma}^{\sigma}F_{\sigma\mu}a^{+\mu}\sum_{k=0}^\infty d_{4k}\,b_1^{+k}b_1^{k+2}
\label{op-L1}
\end{eqnarray}
\begin{eqnarray}
T_0&=&i\tilde{\gamma}^\mu D_\mu-\tilde{\gamma}m
+\tilde{\gamma}^\tau F_{\tau\sigma}D^\sigma \sum_{k=0}^\infty c_{0k}\,b_1^{+k}b_1^k
\nonumber
\\
&&{}
+\tilde{\gamma}a^\alpha F_{\alpha\sigma}D^\sigma \sum_{k=0}^\infty c_{4k}\,(b_1^{+})^{k+1}b_1^k
+\tilde{\gamma}a^{+\mu} F_{\mu\sigma}D^\sigma \sum_{k=0}^\infty c_{5k}\,b_1^{+k}b_1^{k+1}
\nonumber
\\
&&{}
+\tilde{\gamma}\tilde{\gamma}^{\mu\nu}F_{\mu\nu}\sum_{k=0}^\infty a_{0k}\,b_1^{+k}b_1^k
+\tilde{\gamma}a^{+\mu}F_{\mu\alpha}a^\alpha\sum_{k=0}^\infty a_{4k}\,b_1^{+k}b_1^k
\nonumber
\\
&&{}
+\tilde{\gamma}^{\sigma}F_{\sigma\alpha}a^\alpha\sum_{k=0}^\infty a_{2k}\,(b_1^{+})^{k+1}b_1^k
+\tilde{\gamma}^{\sigma}F_{\sigma\mu}a^{+\mu}\sum_{k=0}^\infty a_{3k}\,b_1^{+k}b_1^{k+1}
\label{op-T0}
\end{eqnarray}
\begin{eqnarray}
L_1^+&=&ia^{+\mu}D_\mu+mb_1^+
+a^{+\mu} F_{\mu\sigma}D^\sigma\sum_{k=0}^\infty f_{1k}\,b_1^{+k}b_1^k
+\tilde{\gamma}\tilde{\gamma}^\tau F_{\tau\sigma}D^\sigma\sum_{k=0}^\infty f_{3k}\,(b_1^{+})^{k+1}b_1^k
\nonumber
\\
&&{}
+a^\alpha F_{\alpha\sigma}D^\sigma \sum_{k=0}^\infty f_{5k}\,(b_1^{+})^{k+2}b_1^k
+\tilde{\gamma}^{\mu\nu}F_{\mu\nu}\sum_{k=0}^\infty d_{1k}\,(b_1^{+})^{k+1}b_1^k
\nonumber
\\
&&{}
+\tilde{\gamma}\tilde{\gamma}^{\sigma}F_{\sigma\mu}a^{+\mu}\sum_{k=0}^\infty d_{3k}\,b_1^{+k}b_1^k
\nonumber
\\
&&{}
+a^{+\mu}F_{\mu\alpha}a^\alpha \sum_{k=0}^\infty d_{9k}\, (b_1^{+})^{k+1}b_1^k
+\tilde{\gamma}\tilde{\gamma}^{\sigma}F_{\sigma\alpha}a^\alpha\sum_{k=0}^\infty d_{5k}\,(b_1^{+})^{k+2}b_1^k
\label{op-L1+}
\end{eqnarray}
where $a_{ik}$, $c_{ik}$, $d_{ik}$, $c_{ik}$ are arbitrary complex
constants and the  rest of the operators (\ref{op-t1})--(\ref{op-g0}) are
unchanged as one can see form the equation (\ref{T1}).
Let us note that the above relations can be treated as
the deformations of the corresponding relations of free theory by
the terms linear in $F_{\mu\nu}$.

Let us point out that the ansatz for the operators $L_1$, $T_0$,
$L_1^+$ (\ref{op-L1})--(\ref{op-L1+}) is not the most general one. The
ansatz is taken on the basis  of the following ``minimal" rule. Let us
consider the operators (\ref{op-L1})--(\ref{op-L1+}) in free theory,
replace the partial derivatives by the covariant ones and calculate
the commutators. Obviously the  algebra will not be
closed. Then one adds to these operators the minimal number of terms
linear in $F_{\mu\nu}$ in such a way that  the  algebra is closed in the
linear approximation. One can see that according to this ``minimal"
rule the Lorentz indices of the creation and annihilation operators
are always contracted with the an index or indices of $F_{\mu\nu}$.
In principle it is possible to consider  other deformations of the
free theory by the terms linear in $F_{\mu\nu}$. For example, one can
add to $L_1$ a term of the form $a^\mu \gamma_\mu \gamma^\nu
F_{\nu\sigma}D^\sigma$ but this term  does not obey the ``minimal" rule.

From the requirement the $T_0$ and $L_0$ to be  hermitian, from the
condition $(L_1)^+=L_1^+$ and from the requirement that the total
system of operators forms a closed algebra in the linear
approximation one finds the expressions for constants which are
present in (\ref{op-L1})--(\ref{op-L1+}), These expressions are
summarized in the Appendix.

Note that  a similar problem was considered in
\cite{Klishevich:1998yt}, but we found two more arbitrary constants
because, unlike \cite{Klishevich:1998yt}, we do not require from the
very beginning that the coefficients in
(\ref{op-L1})--(\ref{op-L1+}) must satisfy reality conditions. As one can
see from the Appendix, the complex coefficients are also
acceptable.

The new operators form the algebra which is the same as in the free
case and is given in Table \ref{table0}.
\begin{table}[t]
\small
\begin{eqnarray*}
\begin{array}{||c||c|c|c|c|c|c|c|c||c||}\hline\hline\vphantom{\biggm|}\hspace{-0.3em}
[\;\downarrow\;,\to\}\hspace{-0.4em}&T_0&T_1&T_1^+&\quad L_0&L_1&L_1^+&L_2&L_2^+ &G_0\\
\hline\hline\vphantom{\biggm|}
T_0
   &2L_0&-2L_1&-2L_1^+&0&0&0&0&0&0\\
\hline\vphantom{\biggm|}
T_1
   &-2L_1&-4L_2&-2G_0&0&0&T_0&0&T_1^+&T_1\\
\hline\vphantom{\biggm|}
T_1^+
   &-2L_1^+&-2G_0&-4L_2^+&0&-T_0&0&-T_1&0&-T_1^+ \\
\hline\vphantom{\biggm|}
L_0
   &0&0&0&0&0&0&0&0&0\\
\hline\vphantom{\biggm|}
L_1
   &0&0&T_0&0&0&-L_0&0&L_1^+&L_1 \\
\hline\vphantom{\biggm|}
L_1^+
   &0&-T_0&0&0&L_0&0&-L_1&0&-L_1^+\\
\hline\vphantom{\biggm|}
L_2
   &0&0&T_1&0&0&L_1&0&G_0&2L_2\\
\hline\vphantom{\biggm|}
L_2^+
   &0&-T_1^+&0&0&-L_1^+&0&-G_0&0&-2L_2^+\\
\hline\hline\vphantom{\biggm|}
G_0
   &0&-T_1&T_1^+&0&-L_1&L_1^+&-2L_2&2L_2^+&0\\
\hline\hline
\end{array}
\end{eqnarray*}
\caption{The algebra of the operators.}\label{table0}
\end{table}

After we have achieved the closure of the algebra for the operators, the next step is to construct the corresponding BRST charge.
This procedure follows closely the one developed for the fermionic fields in   \cite{ Buchbinder:2006nu},\cite{Buchbinder:2004gp}
to which we refer for more details.
First we construct the standard BRST operator on the basis of the operators (\ref{T1})--(\ref{op-L1+})
\begin{eqnarray}\label{QQ}
Q&=&q_0T_0+q_1^+T_1+q_1T_1^++\eta_0L_0+\eta_1^+L_1+\eta_1L_1^+
+\eta_2^+L_2+\eta_2 L_2^++\eta_G G_0
\nonumber
\\&&
{}
+2q_0(q_1^+{\cal{}P}_1+q_1{\cal{}P}_1^+)
+(q_1^+\eta_1-\eta_1^+q_1)ip_0
+(\eta_1^+\eta_1-q_0^2){\cal{}P}_0
+2q_1^{+2}{\cal{}P}_2
\nonumber
\\&&
{}+2q_1^2{\cal{}P}_2^+
+q_1^+\eta_2ip_1^+
-\eta_2^+q_1ip_1
-\eta_2^+\eta_1{\cal{}P}_1
-\eta_1^+\eta_2{\cal{}P}_1^+
+(2q_1^+q_1-\eta_2^+\eta_2){\cal{}P}_G
\nonumber
\\&&
{}
+\eta_G(q_1^+ip_1-q_1ip_1^+
+\eta_1^+{\cal{}P}_1-\eta_1{\cal{}P}_1^++2\eta_2^+{\cal{}P}_2-2\eta_2{\cal{}P}_2^+)
\end{eqnarray}
Here, $q_0$, $q_1$, $q_1^+$ and  $\eta_0$, $\eta_1^+$, $\eta_1$,
$\eta_2^+$, $\eta_2$, $\eta_G$ are, respectively, the bosonic and
fermionic ghost ``coordinates'' corresponding to their canonically
conjugate ghost ``momenta'' $p_0$, $p_1^+$, $p_1$, ${\cal{}P}_0$,
${\cal{}P}_1$, ${\cal{}P}_1^+$, ${\cal{}P}_2$, ${\cal{}P}_2^+$,
${\cal{}P}_G$. They obey the (anti)commutation relations
\begin{eqnarray}
\label{ghosts}
&
\{\eta_1,{\cal{}P}_1^+\}= \{{\cal{}P}_1, \eta_1^+\}
=
\{\eta_2,{\cal{}P}_2^+\}= \{{\cal{}P}_2, \eta_2^+\}
=\{\eta_0,{\cal{}P}_0\}= \{\eta_G,{\cal{}P}_G\} =
1,
\nonumber
\\
& [q_0, p_0]=[q_1, p_1^+] = [q_1^+, p_1] = i
\end{eqnarray}
and possess the standard  ghost number distribution,
$gh(q,\eta)$ = $ - gh(p, \mathcal{P})$ = $1$,
which gives    $gh({Q})$ = $1$.

For the subsequent computations it is convenient to present the  BRST operator (\ref{QQ})
in the form
\begin{eqnarray*}
Q&=&
\tilde{Q}+\eta_G\bigl(N+\tfrac{d-3}{2}+h\bigr)+(2q_1^+q_1-\eta_2^+\eta_2){\cal{}P}_G
\nonumber
\\
{}
N&=&a^+_\mu a^\mu+b_1^+b_1+2b_2^+b_2+f^+f
\nonumber
\\
&&{}
+q_1^+ip_1-ip_1^+q_1
+\eta_1^+{\cal{}P}_1+{\cal{}P}_1^+\eta_1+2\eta_2^+{\cal{}P}_2+2{\cal{}P}_2^+\eta_2
\\
[0.5em]
\tilde{Q}&=&
q_0\tilde{T}_0+\eta_0L_0+\Delta Q
+(q_1^+\eta_1-\eta_1^+q_1)ip_0
+(\eta_1^+\eta_1-q_0^2){\cal{}P}_0
\\
[0.5em]
\Delta Q&
=&
q_1^+T_1+q_1T_1^++\eta_1^+L_1+\eta_1L_1^+
+\eta_2^+L_2+\eta_2L_2^+
+2q_1^{+2}{\cal{}P}_2+2q_1^2{\cal{}P}_2^+
\nonumber
\\
&&\qquad{}
+q_1^+\eta_2ip_1^+
-\eta_2^+q_1ip_1
-\eta_2^+\eta_1{\cal{}P}_1
-\eta_1^+\eta_2{\cal{}P}_1^+
\\
\tilde{T}_0
&=&
T_0+2q_1^+{\cal{}P}_1+2q_1{\cal{}P}_1^+
\end{eqnarray*}
Next  we choose the following representation for the vacuum in the  Hilbert space
\begin{equation}
\left( p_0, q_1, p_1, \mathcal{P}_0,
{\cal{}P}_G, \eta_1, {\cal{}P}_1, \eta_2,
{\cal{}P}_2\right)|0\rangle =0\,,
\label{ghostvac}
\end{equation}
and suppose
that the  vectors and gauge parameters do not depend on  $\eta_G$,
\begin{eqnarray} \nonumber \label{chi}
 |\chi\rangle &&=  \\ \nonumber
&&\displaystyle\sum\limits_{k_i}
(q_0)^{k_1}(q_1^+)^{k_2}(p_1^+)^{k_3}(\eta_0)^{k_4}(f^+)^{k_5}
  (\eta_1^+)^{k_6}(\mathcal{P}_1^+)^{k_7}(\eta_2^+)^{k_8} (\mathcal{P}_2^+)^{k_9}
  (b_1^+)^{k_{10}}(b_2^+)^{k_{11}} \times \nonumber \\
   & &\times a^{+{}\mu_1}\cdots a^{+{}\mu_{k_0}}\chi^{k_1 \cdots k_{11}}_{\mu_1\cdots \mu_{k_0}}(x)
   |0\rangle. 
\end{eqnarray}
The sum in (\ref{chi}) is taken  over $k_0, k_1, k_2$, $k_3$,
$k_{10}$, $k_{11}$, running from 0 to infinity, and over $k_4, k_5,
k_6, k_7, k_8, k_9$, running from 0 to 1.
Then, we derive from the
equations (\ref{Q}) as well as from the reducible gauge transformations, (\ref{L}) a sequence of relations
\begin{align}
& \tilde{Q}|\chi\rangle=0, && (N+\tfrac{d-3}{2}+h)|\chi\rangle=0, &&
\left(\epsilon, {gh}\right)(|\chi\rangle)=(1,0),
\label{Qchi}
\\
& \delta|\chi\rangle=\tilde{Q}|\Lambda\rangle, &&
(N+\tfrac{d-3}{2}+h)|\Lambda\rangle=0, && \left(\epsilon,
{gh}\right)(|\Lambda\rangle)=(0,-1),
\label{QLambda}
\\
& \delta|\Lambda\rangle=\tilde{Q}|\Lambda^{(1)}\rangle, &&
(N+\tfrac{d-3}{2}+h)|\Lambda^{(1)}\rangle=0, && \left(\epsilon,
{gh}\right)(|\Lambda^{(1)}\rangle)=(1,-2),\\
& \delta|\Lambda^{(i-1)}\rangle=\tilde{Q}|\Lambda^{(i)}\rangle, &&
(N+\tfrac{d-3}{2}+h)|\Lambda^{(i)}\rangle=0, && \left(\epsilon,
{gh}\right)(|\Lambda^{(i)}\rangle)= (i,-i-1). \label{QLambdai}
\end{align}
Here $\epsilon$ defines a Grassmann parity of  corresponding fields and parameters of gauge transformations
as $(-1)^\epsilon$.

The middle equation in (\ref{Qchi}) is a constraint on   possible values of $h$
\begin{eqnarray}
\label{h}
h=2-s-\frac{d}{2}.
\end{eqnarray}
By fixing the value of spin, we also fix the parameter $h$, according to
(\ref{h}).
Having fixed a value of $h$, we then  substitute it into each of
the expressions (\ref{Qchi})--(\ref{QLambdai}).

Analogously to the free case \cite{Buchbinder:2006nu} the equation
of motion (\ref{Qchi}) cannot be obtained from a Lagrangian. In
order to extract from  (\ref{Qchi}) a Lagrangian set of equations of
motion we decompose the state vector and gauge parameters in terms
of powers of neutral Grassmann even
  $q_0$,
and Grassmann odd
$\eta_0$ ghosts
\begin{eqnarray*}
|\chi\rangle=\sum_{k=0}^\infty q_0^k(|\chi^k_0\rangle+\eta_0|\chi_1^k\rangle),
&\qquad&
|\Lambda\rangle=\sum_{k=0}^\infty q_0^k(|\Lambda^k_0\rangle+\eta_0|\Lambda_1^k\rangle).
\end{eqnarray*}
Then  we remove all fields except $|\chi_0^0\rangle$ and
$|\chi_0^1\rangle$
using
a part of the initial gauge symmetries or using their own equations of
motion.
 As a result of this procedure
the equation  (\ref{Qchi}) is reduced to
\begin{equation}
\Delta Q|\chi_0^0\rangle+\tfrac{1}{2}\{\tilde{T_0},\eta_1^+\eta_1\}|\chi_0^1\rangle=0, \quad
\tilde{T_0}|\chi_0^0\rangle+\Delta Q|\chi_0^1\rangle=0
\label{EofM-0}
\end{equation}
These equations are invariant under the gauge transformations
\begin{equation}
\delta|\chi_0^0\rangle=
\Delta Q|\Lambda_0^0\rangle+\tfrac{1}{2}\{\tilde{T_0},\eta_1^+\eta_1\}|\Lambda_0^1\rangle
\quad
\delta|\chi_0^1\rangle=
\tilde{T_0}|\Lambda_0^0\rangle+\Delta Q|\Lambda_0^1\rangle
\label{GT}
\end{equation}
The parameters of gauge transformations are in turn invariant under
the chain of transformations with a finite number of reducibility
stages $i_{\text{max}}=s-3/2$
\begin{align}
\delta|\Lambda^{(i)}{}^{0}_{0}\rangle
&=
\Delta{}Q|\Lambda^{(i+1)}{}^{0}_{0}\rangle
 +
 \frac{1}{2}\bigl\{\tilde{T}_0,\eta_1^+\eta_1\bigr\}
 |\Lambda^{(i+1)}{}^{1}_{0}\rangle,
&
|\Lambda^{(0)}{}^0_0\rangle_n=|\Lambda^0_0\rangle,
\label{GTi1}
\\
\delta|\Lambda^{(i)}{}^{1}_{0}\rangle
&=
\tilde{T}_0|\Lambda^{(i+1)}{}^{0}_{0}\rangle
 +\Delta{}Q|\Lambda^{(i+1)}{}^{1}_{0}\rangle,
&
|\Lambda^{(0)}{}^1_0\rangle_n=|\Lambda^1_0\rangle,
\label{GTi2}
\\
&
i_{\text{max}}=s-3/2
\label{imax}
\end{align}
where $\{\tilde{T_0},\eta_1^+\eta_1\}=\tilde{T_0}\eta_1^+\eta_1+\eta_1^+\eta_1\tilde{T_0}.$

It is straightforward to check that the equations (\ref{EofM-0}) can be obtained from the following Lagrangian
\begin{eqnarray}
\mathcal{L}&=&
\langle\tilde{\chi}^0_0|K_h\Bigl\{\tilde{T_0}|\chi_0^0\rangle+\Delta Q|\chi_0^1\rangle\Bigr\}
+\langle\tilde{\chi}^1_0|K_h\Bigl\{\Delta Q|\chi_0^0\rangle+\tfrac{1}{2}\{\tilde{T_0},\eta_1^+\eta_1\}|\chi_0^1\rangle\Bigr\}
,
\label{LagrF}
\end{eqnarray}
In (\ref{LagrF}) operator $K_h$
\begin{eqnarray}
K_h&=&\sum_{n=0}^\infty\frac{1}{n!}\Bigl(|n\rangle\langle n|C(n,h)-2f^+|n\rangle\langle n| f\, C(n+1,h)\;\Bigr),
\label{K}
\\
&&
C(n,h)=h(h+1)\cdots(h+n-1),
\qquad
C(0,h)=1,
\qquad
|n\rangle=(b_2^+)^n|0\rangle
\nonumber
\end{eqnarray}
is
needed to maintain hermiticity of the Lagrangian since
 as one can see from the auxiliary representations for operators
(\ref{op-t1}) -- (\ref{op-t1}) one has
$(l_2)^+ \neq l_2^+$ and $(t_1)^+\neq t_1^+$.
The fields
 $\langle\tilde{\chi}^0_0|$, $\langle\tilde{\chi}^1_0|$ are defined as follows
\begin{eqnarray}
\langle\tilde{\chi}^0_0|=(|\chi_0^0\rangle)^+\tilde{\gamma}^0,
&\qquad&
\langle\tilde{\chi}^1_0|=(|\chi_0^1\rangle)^+\tilde{\gamma}^0.
\end{eqnarray}

The Lagrangian (\ref{LagrF}) describes  the interaction of massive fermionic fields with constant electromagnetic field and it is
 our main result.
It contains, apart from the physical field $\psi_{\mu_1\ldots\mu_n}(x)$ in $|\chi_0^0\rangle$
\begin{eqnarray}
|\chi_0^0\rangle=\psi_{\mu_1\ldots\mu_n}(x)a^{+\mu_1}\ldots a^{+\mu_n}|0\rangle
+\ldots
\end{eqnarray}
a number of auxiliary fields\footnote{In decomposition (\ref{chi}) they are coefficients in summands which contain at least one creation operator different from $a^{+\mu}$.}, whose number increases with spin value.
One can partially or completely fix gauge invariance and
obtain different Lagrangian formulations with  a smaller number of auxiliary
fields, as we shall do it in the next Section.

\setcounter{equation}0\section{Lagrangian formulations with  a smaller number of auxiliary
fields} \label{secquartet}

In this Section we are going to obtain from (\ref{LagrF}) different Lagrangian formulations partially fixing the gauge invariance.

First we derive a quartet Lagrangian formulation  \cite{Buchbinder:2007ak,Buchbinder:2008ss}.
Initially this formulation was developed for the massless higher spin fields in flat and AdS background in \cite{Buchbinder:2007ak}. Its fermionic version  contains seven unconstrained fields (one physical field and six auxiliary fields three of which are Lagrangian multipliers) and one unconstrained gauge parameter.\footnote{Another similar formulation (so-called triplet formulation) of fermionic fields
on Minkowski and $AdS_d$ backgrounds contains one physical and two
auxiliary fields \cite{Francia:2002pt}--\cite{Sorokin:2008tf} (see also \cite{Bekaert:2015fwa} for a recent discussion) and
corresponds to a description of reducible representations of the
Poincare or $SO(d-2,2)$ groups.}
 Using  dimensional reduction one can obtain the quartet formulation  for massive higher spin fields in Minkowski space  \cite{Buchbinder:2008ss}.

To obtain this formulation from the Lagrangian (\ref{LagrF}) we partially fix gauge invariance just as it was done in \cite{Buchbinder:2007vq}, except we will not fix gauge invariance corresponding to gauge parameter $|\varepsilon\rangle$
\begin{eqnarray}
|\Lambda^{(0)}{}^0_{0}\rangle
&=&
\mathcal{P}_1^+|\varepsilon\rangle
+\ldots
\qquad
|\varepsilon\rangle=\sum_{k=0}^{n-1}\frac{1}{k!}(b_1^+)^k|\varepsilon_{n-k-1}\rangle
\\
&&
|\varepsilon_{n-k-1}\rangle=\frac{1}{(n-k)!}\;a^{+\mu_1}\ldots a^{+\mu_{n-k-1}}\varepsilon_{\mu_1\ldots\mu_{n-k-1}}(x)|0\rangle
.
\end{eqnarray}
Next one can show that after the gauge fixing some of the remaining fields can be removed with the help of the equations of motion and the nonvanishing fields in the quartet formulation are
\begin{eqnarray}
|\chi_0^0\rangle&=&|\Psi^{(n)}\rangle+\eta_1^+\mathcal{P}_1^+|D^{(n-2)}\rangle+q_1^+\mathcal{P}_1^+|E^{(n-2)}\rangle+i\eta_1^+p_1^+|\Sigma^{(n-2)}\rangle
\\
|\chi_0^1\rangle&=&\mathcal{P}_1^+|C^{(n-1)}\rangle-ip_1^+|\Lambda^{(n-1)}\rangle+ip_1^+\eta_1^+\mathcal{P}_1^+|\Omega^{(n-3)}\rangle
\end{eqnarray}

The Lagrangian and the gauge transformation
for the massive fermionic higher spin field interacting with constant electromagnetic field
in the quartet formulation
are\footnote{
In order to obtain triplet formulation \cite{Francia:2002pt}--\cite{Sorokin:2008tf} one should to discard field $|E^{(n-2)}\rangle$ and Lagrangian multipliers
$|\Lambda^{(n-1)}\rangle$, $|\Sigma^{(n-2)}\rangle$, $|\Omega^{(n-3)}\rangle$ in (\ref{Lagr-q}).}
\begin{eqnarray}
{\cal{}L}
&=&\langle\tilde{\Psi}^{(n)}|
\Bigl\{
T_0|\Psi^{(n)}\rangle+L_1^+|C^{(n-1)}\rangle+T_1^{\prime+}|\Lambda^{(n-1)}\rangle
\Bigr\}
\nonumber
\\
&&{}
-\langle\tilde{C}^{(n-1)}|
\Bigl\{
T_0|C^{(n-1)}\rangle-L_1|\Psi^{(n)}\rangle+L_1^+|D^{(n-2)}\rangle-|\Lambda^{(n-1)}\rangle-T_1^{\prime+}|\Sigma^{(n-2)}\rangle
\Bigr\}
\nonumber
\\
&&{}
-\langle\tilde{D}^{(n-2)}|
\Bigl\{
T_0|D^{(n-2)}\rangle+L_1|C^{(n-1)}\rangle+2|\Sigma^{(n-2)}\rangle-T_1^{\prime+}|\Omega^{(n-3)}\rangle
\Bigr\}
\nonumber
\\
&&{}
+\langle\tilde{\Lambda}^{(n-1)}|
\Bigl\{
T_1'|\Psi^{(n)}\rangle+|C^{(n-1)}\rangle+L_1^+|E^{(n-2)}\rangle
\Bigr\}
\nonumber
\\
&&{}
+\langle\tilde{\Sigma}^{(n-2)}|
\Bigl\{
T_1'|C^{(n-1)}\rangle-2|D^{(n-2)}\rangle
+T_0|E^{(n-2)}\rangle
\Bigr\}
\nonumber
\\
&&{}
+\langle\tilde{\Omega}^{(n-3)}|
\Bigl\{
T_1'|D^{(n-2)}\rangle+L_1|E^{(n-2)}\rangle
\Bigr\}
\nonumber
\\
&&{}
+\langle\tilde{E}^{(n-2)}|
\Bigl\{
L_1|\Lambda^{(n-1)}\rangle+T_0|\Sigma^{(n-2)}\rangle+L_1^+|\Omega^{(n-3)}\rangle
\Bigr\}
\label{Lagr-q}
\end{eqnarray}
\begin{align}
&
\delta|\Psi^{(n)}\rangle=L_1^+|\Upsilon^{(n-1)}\rangle
,
&&
\delta|C^{(n-1)}\rangle=-T_0|\Upsilon^{(n-1)}\rangle
,
\\
&
\delta|D^{(n-2)}\rangle=L_1|\Upsilon^{(n-1)}\rangle
,
&&
\delta|E^{(n-2)}\rangle=-T_1'|\Upsilon^{(n-1)}\rangle
\label{GT-q}
\end{align}
The fields and the gauge parameter $|\Upsilon^{(n-1)}\rangle\equiv|\varepsilon^{(n-1)}\rangle$
depend only on the oscillators
 $(a_\mu,b_1)$.
In particular in  (\ref{Lagr-q})--(\ref{GT-q}) the fields and the gauge parameter have uniform decomposition
\begin{eqnarray}
|\Phi^{(m)}\rangle&=&\sum_{k=0}^m\frac{1}{k!}(b_1^+)^k|\phi_{m-k}\rangle
\\
&&
|\phi_{m-k}\rangle=\frac{1}{(n-k)!}\;a^{+\mu_1}\ldots a^{+\mu_{m-k}}\phi_{\mu_1\ldots\mu_{m-k}}(x)|0\rangle
\label{decomp-q}
\end{eqnarray}
and the operators $T_1'$ and $T_1^{\prime+}$ are the $(a_\mu,b_1)$ parts of the operators $t_1$ and $t_1^+$ (\ref{op-t1})
\begin{align}
&
T_1'=\tilde{\gamma}^\mu a_\mu-\tilde{\gamma}b_1
&&
T_1^{\prime+}= a_\mu^+\tilde{\gamma}^\mu-\tilde{\gamma}b_1^+
.
\end{align}

Next we will show that the Lagrangian formulation, which obtained in
\cite{Klishevich:1998yt}, is a particular case of our general result
(\ref{Lagr-q}). To get such Lagrangian formulations we first partly
fix the gauge, removing the field $|E^{(n-2)}\rangle$ with the help of
gauge transformations (\ref{GT-q}) and then integrate out all the
fields except the field $|\Psi^{(n)}\rangle$. The result is
\begin{eqnarray}
{\cal{}L}
&=&\langle\tilde{\Psi}^{(n)}|
\Bigl\{
T_0-L_1^+T_1'-T_1^{\prime+}L_1
-T_1^{\prime+}T_0T_1'
\nonumber
\\
&&\qquad\qquad{}
-\tfrac{1}{2}T_1^{\prime+}L_1^+T_1'T_1'-\tfrac{1}{2}T_1^{\prime+}T_1^{\prime+}L_1T_1'
-\tfrac{1}{4}T_1^{\prime+}T_1^{\prime+}T_0T_1'T_1'
\Bigr\}|\Psi^{(n)}\rangle
\label{Lagr-3}
\\
&&
\delta|\Psi^{(n)}\rangle=L_1^+|\Upsilon^{(n-1)}\rangle
\qquad
\label{GT-q1}
\end{eqnarray}
where the state
 $|\Psi^{(n)}\rangle$ and the parameter of gauge transformations $|\Upsilon^{(n-1)}\rangle$ obey the constraints
\begin{align}
&
(T_1')^3|\Psi^{(n)}\rangle=0,
&&
T_1'|\Upsilon^{(n-1)}\rangle=0.
\label{triple}
\end{align}
Such a partial form of the Lagrangian was obtained in
\cite{Klishevich:1998yt}, but with another (less general\footnote{It
should be noted that in \cite{Klishevich:1998yt} was considered
deformation of the operators corresponding to the gamma-traceless
conditions as well. But this deformation is proportional to an
arbitrary constant and as we said at the beginning of our paper can
be removed by a field redefinition.}) expressions for the operators
(\ref{T1})--(\ref{op-L1+}).

We can proceed to obtain more Lagrangian formulations.
For example, we can resolve constraints on the field and the gauge parameter (\ref{triple}).
Using decomposition (\ref{decomp-q}) for $|\Psi^{(n)}\rangle$ and $|\Upsilon^{(n-1)}\rangle$
\begin{eqnarray}
\label{res-psi}
|\Psi^{(n)}\rangle=\sum_{k=0}^{n}\frac{1}{k!}(b_1^+)^{k}|\psi_{n-k}\rangle
&\qquad&
|\Upsilon^{(n-1)}\rangle=\sum_{k=0}^{n-1}\frac{1}{k!}(b_1^+)^{k}|\epsilon_{n-1-k}\rangle
\end{eqnarray}
we find that gauge parameter $|\epsilon_{n-1}\rangle$ is not restricted and the other parameters $|\epsilon_{k}\rangle$ are expressed in terms of its gamma-traces
$|\epsilon_{k}\rangle=(\gamma^\mu a_\mu)^{n-1-k}|\epsilon_{n-1}\rangle,$
so we may make gauge transformation using the unrestricted gauge parameter $|\epsilon_{n-1}\rangle.$
One can do the same  for the field  $|\Psi^{(n)}\rangle$. Due to restriction (\ref{triple}) there are only three independent fields $|\psi_{n}\rangle$, $|\psi_{n-1}\rangle$, $|\psi_{n-2}\rangle$ and all the other fields are expressed through these three fields
\begin{align}
&|\psi_{n-2k-1}\rangle=-k(\gamma^\mu a_\mu)^{2k+1}|\psi_{n}\rangle
+(\gamma^\mu a_\mu)^{2k}|\psi_{n-1}\rangle+(k+1)(\gamma^\mu a_\mu)^{2k-1}|\psi_{n-2}\rangle
\\ \nonumber
&k\ge1, \\ 
&|\psi_{n-2k-2}\rangle=-k(\gamma^\mu a_\mu)^{2k+2}|\psi_{n}\rangle
+(k+1)(\gamma^\mu a_\mu)^{2k}|\psi_{n-2}\rangle
.
\end{align}

Thus one can obtain\footnote{Since the Lagrangian formulation is
very large, we do not present it here.} a gauge invariant Lagrangian
formulation for a  massive fermionic field interacting with constant
electromagnetic field with the help of three fields
$|\psi_{n}\rangle$, $|\psi_{n-1}\rangle$, $|\psi_{n-2}\rangle$ and
one gauge parameter $|\epsilon_{n-1}\rangle$.

Finally,
using the remaining unrestricted gauge parameter $|\epsilon_{n-1}\rangle$ one can remove field $|\psi_{n-1}\rangle$ and obtain
 a Lagrangian formulation in terms of two traceful unrestricted fields: one physical $|\psi_{n}\rangle$ field
 and one auxiliary $|\psi_{n-2}\rangle$ field. This Lagrangian  has no gauge invariance since
we have already used  entire gauge freedom. It should be noted that
if we decompose the tracefull fields $|\psi_{n}\rangle$ and
$|\psi_{n-2}\rangle$ in a series of traceless fields we obtain set
of the fields which coincide with the set of fields of Singh and
Hagen \cite{SH}.

\setcounter{equation}0\section{Example: spin 3/2}\label{ex-32}

In this section we apply a general procedure described in the previous Sections for the simplest example
of spin-3/2 field.

In the case of spin-3/2 field we have $h=\frac{1-d}{2}$ (see eq.
(\ref{h})) and since according to (\ref{imax}) we have  $i_{max}=0$.
Therefore the corresponding Lagrangian formulation is an
irreducible gauge theory. Due to ${gh}(|\Lambda^1_0\rangle_1 =
-2$), the nonvanishing fields $|\chi^0_0\rangle_1$,
$|\chi^1_0\rangle_1$ and the gauge parameter
$|\Lambda^0_0\rangle_1$, (we have $|\Lambda^1_0\rangle_1 \equiv 0$),
possess the following Grassmann grading and ghost number
distributions:
\begin{eqnarray}
\left(\varepsilon, {gh}\right)(|\chi_0^0\rangle_1)=(1, 0),\quad
\left(\varepsilon, {gh}\right)(|\chi_0^1\rangle_1)=(1, -1),\quad
\left(\varepsilon, {gh}\right)(|\Lambda_0^0\rangle_1)=(0,-1).
\end{eqnarray}
These conditions determine the dependence of the fields and of the gauge
parameters on the oscillator variables in a unique form
\begin{eqnarray*}
|\chi^0_0\rangle_1 &=&
\bigl[ia^{+\mu}\psi_\mu(x)+f^+\tilde{\gamma}\psi(x)
  +b_1^+\varphi(x)\bigr]
|0\rangle
,
\\
\langle\tilde{\chi}{}^0_0|
&=&\langle0|\bigl[
-\psi_\mu^+(x)ia^{\mu}+\psi^+(x)\tilde{\gamma}f
+\varphi^+(x)b_1
\bigr]\tilde{\gamma}{}^0
,
\\
|\chi^1_0\rangle_1
&=&
\bigl[{\cal{}P}_1^+\tilde{\gamma}\chi(x)
 -ip_1^+\chi_1(x)\bigr]|0\rangle
,
\\
\langle\tilde{\chi}{}^1_0|
&=&
\langle0|\bigl[\chi_1^+(x)ip_1
+\chi^+(x)\tilde{\gamma}{\cal{}P}_1\Bigr]\tilde{\gamma}{}^0
,
\\
|\Lambda^0_0\rangle_1
&=&
\bigl[{\cal{}P}_1^+\lambda(x)
 -ip_1^+\tilde{\gamma}\lambda_1(x)\bigr]|0\rangle
\qquad
h=-\tfrac{d-1}{2}
.
\end{eqnarray*}
Substituting these expressions for the fields and the gauge
parameters in (\ref{LagrF}) and (\ref{GT}) one finds the  Lagrangian
and gauge transformations for the physical spin-3/2 field $\psi^\mu$
and for the auxiliary fields
\begin{eqnarray}\label{L32}
{\cal{}L}_{3/2}&&= \\ \nonumber
&&\bar{\psi}{}^\mu\Bigl[
(i\gamma^\sigma D_\sigma-m)\psi_\mu
+\frac{ie}{2m^2}(2\zeta_1+\xi_1)\gamma^\tau F_{\tau\sigma}D^\sigma \psi_\mu
-\frac{ie}{8m}(1+4\zeta_0)\gamma^{\alpha\beta}F_{\alpha\beta}\psi_\mu
\\ \nonumber
&&
+\frac{e}{m^2}(\xi_1-2i\zeta_0)F^{\mu\sigma}D_\sigma\varphi
-\frac{e}{2m}(1-2i\xi_1)\gamma^\tau F_{\tau\mu}\varphi
-\frac{ie}{2m}(1-4\zeta_0)F_{\mu\nu}\psi^\nu
\\ \nonumber
&&+D_\mu\chi
+\frac{e}{m^2}(\zeta_1+i\zeta_0)F_{\mu\sigma}D^\sigma\chi 
-\frac{e}{4m}(1+2i\xi_1)\gamma^\tau F_{\tau\mu}\chi
-i\gamma_\mu\chi_1
\Bigr]
\\
\nonumber
&&+
\bar{\varphi}\Bigl[
(i\gamma^\mu D_\mu-m)\varphi
+\frac{ie}{2m^2}(2\zeta_1-3\xi_1)\gamma^\tau F_{\tau\sigma}D^\sigma\varphi
-\frac{e}{m^2}(\xi_1+2i\zeta_0)F^{\mu\sigma}D_\sigma\psi_\mu
\\ \nonumber
&&
-\frac{e}{2m}(1+2i\xi_1)\gamma_\tau F^{\tau\mu}\psi_\mu
-\frac{ie}{8m}(1+4\zeta_0)\gamma^{\mu\nu}F_{\mu\nu}\varphi
\\ \nonumber
&&
+m\chi
+\frac{e}{2m^2}(2\zeta_0+i\xi_1)\gamma^\tau F_{\tau\sigma}D^\sigma\chi
-\dfrac{ie}{8m}(1-4\zeta_0+4i\xi_1)\gamma^{\mu\nu}F_{\mu\nu}\chi
-\chi_1
\Bigr]
\nonumber
\\ \nonumber
&&
-(d-1)\bar{\psi}\Bigl[
(i\gamma^\mu D_\mu+m)\psi
+\frac{ie}{2m^2}(2\zeta_1+\xi_1)\gamma^\tau F_{\tau\sigma}D^\sigma\psi
\\ \nonumber
&&+\frac{ie}{8m}(1+4\zeta_0)\gamma^{\mu\nu}F_{\mu\nu}\psi
-\chi_1
\Bigr]
\\ \nonumber
&&-
\bar{\chi}\Bigl[
\Bigl\{i\gamma^\mu D_\mu+m+\frac{ie}{2m^2}(2\zeta_1+\xi_1)\gamma^\tau F_{\tau\sigma}D^\sigma
+\frac{ie}{8m}(1+4\zeta_0)\gamma^{\mu\nu}F_{\mu\nu}
\Bigr\}\chi
-\chi_1
\\ \nonumber
&&
+\Bigl\{D_\mu+\frac{e}{m^2}(\zeta_1-i\zeta_0)F_{\mu\sigma}D^\sigma
+\frac{e}{4m}(1-2i\xi_1)\gamma^\tau F_{\tau\mu}
\Bigr\}\psi^\mu
\\ \nonumber
&&
-m\varphi
+\frac{e}{2m^2}(2\zeta_0-i\xi_1)\gamma^\tau F_{\tau\sigma}D^\sigma\varphi
+\frac{ie}{8m}(1-4\zeta_0-4i\xi_1)\gamma^{\mu\nu}F_{\mu\nu}\varphi
\Bigr]
\nonumber
\\
&&{}
+\bar{\chi}_1\Bigl[
\chi+i\gamma^\mu\psi_\mu-\varphi+(d-1)\psi
\Bigr] \nonumber
\end{eqnarray}

\begin{eqnarray}\label{G32}
&&\delta \psi_\mu=\Bigl\{D_\mu+\frac{e}{m^2}(\zeta_1+i\zeta_0)F_{\mu\sigma}D^\sigma
+\frac{e}{4m}(1+2i\xi_1)\gamma^\sigma F_{\sigma\mu}\Bigr\}\lambda
-i\gamma_\mu\lambda_1,
\\ \nonumber
&&
\delta \psi=\lambda_1,
\\ \nonumber
&&
\delta\varphi=\Bigl\{
m-\dfrac{e}{2m^2}\bigl(2\zeta_0+i\xi_1\bigr)\gamma^\tau F_{\tau\sigma}D^\sigma
-\dfrac{ie}{8m}(1-4\zeta_0+4i\xi_1)\gamma^{\mu\nu}F_{\mu\nu}
\Bigr\}\lambda
+\lambda_1,
\\ \nonumber
&&
\delta\chi
=\Bigl\{
-i\gamma^\mu D_\mu+m-\frac{ie}{2m^2}(2\zeta_1+\xi_1)\gamma^\tau F_{\tau\sigma}D^\sigma
+\frac{ie}{8m}(1+4\zeta_0)\gamma^{\mu\nu}F_{\mu\nu}
\Bigr\}\lambda
+2\lambda_1,
\\ \nonumber
&&
\delta \chi_1=\Bigl\{i\gamma^\mu D_\mu+m+\frac{ie}{2m^2}(2\zeta_1+\xi_1)\gamma^\tau F_{\tau\sigma}D^\sigma
+\frac{ie}{8m}(1+4\zeta_0)\gamma^{\mu\nu}F_{\mu\nu}
\Bigr\}\lambda_1.
\end{eqnarray}
Here we have used that $K_hf^+|0\rangle=-2hf^+|0\rangle$ with substitution $-2h\to(d-1)$.

Thus we have derived from the general Lagrangian the one which
contains component fields and the corresponding gauge
transformations. This Lagrangian describes a massive field with spin
$3/2$, coupled to a constant electromagnetic background in the
linear approximation and contains a number of  free
parameters
\footnote{The problem of Lagrangian formulation
for spin- $3/2$ field coupled to EM field  in a linear
approximation has been studied in \cite{Buchbinder:2014} where the
Lagrangian also contains a number of free parameters. However, unlike our
paper,  it is has been assumed  in  \cite{Buchbinder:2014} that the
electromagnetic field is dynamical and moreover,  the model under
consideration possesses a certain ammount  of supersymmetries.
These requirements   impose  the some strong restrictions
on the  structure of the Lagrangian. As a result, the
Lagrangian (\ref{L32}) contains more free parameters in
comparison with  the Lagrangian given in \cite{Buchbinder:2014}.}.
 The
relations (\ref{L32}) -- (\ref{G32}) are our final results. One can
further eliminate the auxiliary fields using the gauge freedom and
some of the equations of motion and thus obtain the field equations
for only physical field $\psi^\mu$.

\section{Conclusions}
In the present paper we have developed the BRST approach to
construct and analyze a Lagrangian description of massive higher
spin fermionic fields interacting with constant electromagnetic
field in the linear approximation. To this end, we  modified  the
operators underlying the BRST charge which corresponds to the
noninteracting fermionic massive higher spin fields by terms
depending on the electromagnetic field. The obtained Lagrangian
contains the auxiliary St\"{u}ckelberg fields which provide the
gauge invariant description for massive theory,  and the  number of these
fields grows with the value of the spin.

We also showed  that one can partially or completely fix the gauge
invariance and obtain a family of  different Lagrangian
formulations with a smaller number of auxiliary fields. As an
example we derived a Lagrangian formulation for the massive
fermionic higher spin fields interacting with a constant
electromagnetic background in the quartet formulation
\cite{Buchbinder:2007ak,Buchbinder:2008ss} and obtained the results
of paper \cite{Klishevich:1998yt} as a  particular case. Also we
gave a detailed description of the component Lagrangian and gauge
transformations for a simplest example of the spin $\frac{3}{2}$
field interacting with a constant electromagnetic background.

Since in the present paper we have considered fermionic higher spin
fields it would be naturally interesting  to generalize the present
results for the case of supersymmetric systems\footnote{Lagrangian
formulation of free supersymmetric massive higher spin theory was
done in \cite{Z}.} as well as to consider higher order interactions.
Inclusion of a nontrivial gravitational background is yet another
interesting problem to consider (see for example
\cite{Florakis:2014kfa}--\cite{Bekaert:2014cea} for recent progress
in these directions). It would be interesting also to establish more
connection with the recent studies in conformal higher spin fields
(see for example \cite{Metsaev:2009ym}--\cite{Metsaev:2014sfa}). We
hope to address these questions in  future publications.

\section*{Acknowledgments}
I.L.B and V.A.K are grateful to the grant for LRSS, project
88.2014.2 and RFBR grants, projects No 13-02-90430 and No
15-02-03594 for partial support. Research of I.L.B was also
supported in part by Russian Ministry of Education and Science under
contract No 2014/387, grant 122. The work of M.T. has been supported
in part by an Australian Research Council grant DP120101340. M.T.
would also like to acknowledge grant 31/89 of the Rustaveli National
Science Foundation.

\setcounter{equation}0
\appendix
\numberwithin{equation}{section}

\section{Expressions for free parameters}\label{Appendix A}
Below we give the expressions for free parameters which are present in the equations (\ref{op-L1})--(\ref{op-L1+})
\begin{align*}
&a_{0(0)}=-\frac{ie}{8m}-\frac{ie}{2m}\zeta_0
&&a_{0(k)}=0\quad k\ge1
\\
&a_{2(0)}=\frac{ie}{2m}-\frac{e}{m}\xi_1
&&a_{2(k)}=-\frac{(-2)^k}{k!}\frac{e}{m}\xi_1\quad k\ge1
\\
&a_{3(0)}=-\frac{ie}{2m}-\frac{e}{m}\xi_1
&&a_{3(k)}=-\frac{(-2)^k}{k!}\frac{e}{m}\xi_1\quad k\ge1
\\
&a_{4(0)}=-\frac{ie}{2m}+\frac{2ie}{m}\zeta_0
&&a_{4(k)}=0\quad k\ge1
\end{align*}
\begin{align*}
&
c_{0(0)}=\frac{ie}{2m^2}(2\zeta_1+\xi_1)
&&c_{0(k)}=\frac{(-2)^k}{k!}\frac{ie}{m^2}\xi_1\quad k\ge1
\\
&c_{4(0)}=\frac{e}{m^2}(-2\zeta_0+i\xi_1)
&&c_{4(k)}=\frac{(-2)^k}{k!}\frac{ie}{m^2}\xi_1\quad k\ge1
\\
&c_{5(0)}=\frac{e}{m^2}(2\zeta_0+i\xi_1)
&&c_{5(k)}=\frac{(-2)^k}{k!}\frac{ie}{m^2}\xi_1\quad k\ge1
\end{align*}
\begin{align*}
&d_{0(0)}=-\frac{ie}{8m}+\frac{ie}{2m}(\zeta_0+i\xi_1)
&&d_{0(k)}=\frac{(-2)^{k-1}}{k!}\frac{e}{m}\xi_1\quad k\ge1
\\
&
d_{1(0)}=-\frac{ie}{8m}+\frac{ie}{2m}(\zeta_0-i\xi_1)
&&d_{1(k)}=-\frac{(-2)^{k-1}}{k!}\frac{e}{m}\xi_1\quad k\ge1
\\
&d_{2(0)}=-\frac{ie}{4m}-\frac{e}{2m}\xi_1
&&d_{2(k)}=\frac{(-2)^{k-1}}{k!}(k+1)\frac{e}{m}\xi_1\quad k\ge1
\\
&d_{3(0)}=-\frac{ie}{4m}+\frac{e}{2m}\xi_1
&&d_{3(k)}=-\frac{(-2)^{k-1}}{k!}(k+1)\frac{e}{m}\xi_1\quad k\ge1
\end{align*}
\begin{align*}
&
d_{4(k)}=\frac{(-2)^k}{k!}\frac{e}{m}\xi_1\quad k\ge0
\\
&d_{5(k)}=-\frac{(-2)^k}{k!}\frac{e}{m}\xi_1\quad k\ge0
\\
&d_{8(0)}=\frac{ie}{m}-\frac{ie}{m}(2\zeta_0+i\xi_1)
&&d_{8(k)}=\frac{(-2)^k}{k!}\frac{e}{m}\xi_1\quad k\ge1
\\
&d_{9(0)}=\frac{ie}{m}-\frac{ie}{m}(2\zeta_0-i\xi_1)
&&d_{9(k)}=-\frac{(-2)^k}{k!}\frac{e}{m}\xi_1\quad k\ge1
\end{align*}
\begin{align*}
&
f_{0(0)}=\frac{e}{m^2}(\zeta_0+i\zeta_1)
\\
&
f_{0(1)}=\frac{e}{m^2}(2\zeta_0-i\xi_1)
&&f_{0(k)}=-\frac{(-2)^{k-1}}{k!}\frac{ie}{m^2}\xi_1\quad k\ge2
\\
&f_{1(0)}=\frac{e}{m^2}(-\zeta_0+i\zeta_1)
\\
&
f_{1(1)}=-\frac{e}{m^2}(2\zeta_0+i\xi_1)
&&f_{1(k)}=-\frac{(-2)^{k-1}}{k!}\frac{ie}{m^2}\xi_1\quad k\ge2
\end{align*}
\begin{align*}
&f_{2(0)}=\frac{e}{m^2}(\zeta_0-\frac{i}{2}\xi_1)
&&f_{2(k)}=\frac{(-2)^{k-1}}{k!}\frac{ie}{m^2}\xi_1\quad k\ge1
\\
&f_{3(0)}=\frac{e}{m^2}(\zeta_0+\frac{i}{2}\xi_1)
&&f_{3(k)}=-\frac{(-2)^{k-1}}{k!}\frac{ie}{m^2}\xi_1\quad k\ge1
\\
&f_{4(0)}=-\frac{2e}{m^2}\zeta_0
&&f_{4(k)}=0\quad k\ge1
\\
&f_{5(0)}=\frac{2e}{m^2}\zeta_0
&&f_{5(k)}=0\quad k\ge1
\end{align*}
Here $\zeta_0$, $\zeta_1$, $\xi_1$ are arbitrary real dimensionless constants.

\end{document}